\documentclass[12pt]{article}
\usepackage{amsmath}
\usepackage{graphicx,psfrag,epsf}
\usepackage{enumerate}
\usepackage{natbib}
\usepackage{textcomp}
\usepackage[hyphens]{url} 
\usepackage{hyperref}

\newcommand{\blind}{0}

\usepackage{longtable,booktabs,array}
\usepackage{calc} 
\usepackage{etoolbox}
\makeatletter
\patchcmd\longtable{\par}{\if@noskipsec\mbox{}\fi\par}{}{}
\makeatother
\IfFileExists{footnotehyper.sty}{\usepackage{footnotehyper}}{\usepackage{footnote}}
\makesavenoteenv{longtable}

\usepackage{graphics}
\usepackage{float}
\usepackage{booktabs}
\usepackage{longtable}
\usepackage{array}
\usepackage{multirow}
\usepackage{wrapfig}
\usepackage{colortbl}
\usepackage{pdflscape}
\usepackage{tabu}
\usepackage{threeparttable}
\usepackage{threeparttablex}
\usepackage[normalem]{ulem}
\usepackage{makecell}
\usepackage{xcolor}

\begin{document}

\def\spacingset#1{\renewcommand{\baselinestretch}%
{#1}\small\normalsize} \spacingset{1}


\if0\blind
{
  \title{\bf The Current State of Undergraduate Bayesian Education and Recommendations for the Future}

  \author{
        Mine Dogucu \thanks{Dogucu was supported by NSF grant IIS-2123366. The authors gratefully acknowledge data collection support of Feiyi Sun.} \\
    Department of Statistics, University of California Irvine\\
     and \\     Jingchen Hu \\
    Department of Mathematics and Statistics, Vassar College\\
      }
  \maketitle
} \fi

\if1\blind
{
  \bigskip
  \bigskip
  \bigskip
  \begin{center}
    {\LARGE\bf The Current State of Undergraduate Bayesian Education and Recommendations for the Future}
  \end{center}
  \medskip
} \fi

\bigskip
\begin{abstract}
As a result of the increased emphasis on mis- and over-use of \(p\)-values in scientific research and the rise in popularity of Bayesian statistics, Bayesian education is becoming more important at the undergraduate level.
With the advances in computing tools, Bayesian statistics is also becoming more accessible for the undergraduates.
This study focuses on analyzing Bayesian courses for the undergraduates.
We explored whether an undergraduate Bayesian course is offered in our sample of 152 high-ranking research universities and liberal arts colleges.
For each identified Bayesian course, we examined how it fits into the institution's undergraduate curricula, such as majors and prerequisites.
Through a series of course syllabi analyses, we explored the topics covered and their popularity in these courses, and the adopted teaching and learning tools, such as software.
This paper presents our findings on the current practices of teaching full Bayesian courses at the undergraduate level.
Based on our findings, we provide recommendations for programs that may consider offering Bayesian courses to their students.
\end{abstract}

\noindent%
\vfill

\newpage
\spacingset{1.45} 

\hypertarget{intro}{%
\section{Introduction}\label{intro}}

In the American Statistical Association's Statement on \(p\)-values, George Cobb's comment on the ASA Discussion Forum highlighted how the use of \(p\)-value of 0.05 is the norm because colleges and graduate schools teach it, and subsequently the scientific community uses it.
On the other hand, the scientific community uses it because colleges and graduate schools teach it \citep{wasserstein2016}.
Evidently, educational programs are at the heart of scientific practice.
In addition to frequentist methods where \(p\)-value is at the core of statistical inference, Bayesian methods provide a different and perhaps more intuitive perspective.
A recent preprint advocates for advancing a Bayesian perspective on probability and uncertainty in science education, with the goal of building trust in science \citep{Rosenberg2022}.
Thanks to recent and ongoing developments in probabilistic programming, Bayesian statistics is becoming more accessible to practitioners and therefore more widely used.
In order for the scientific community to benefit from its advances, Bayesian statistics needs to find its place in educational programs.

Bayesian statistics as a course is more often taught at the graduate level, if taught at all.
As such, students, undergraduate and graduate alike, miss the opportunity to learn Bayesian methods for statistical inference as part of their education training.
As Bayesian methods are more widely used in research and applications nowadays, providing such training for undergraduates is becoming ever more useful and urgent.
Moreover, many Bayesian concepts may help undergraduate students approach scientific inquiries intuitively.
For instance, many misinterpret the \(p\)-value as a hypothesis being true given the data \citep{sotos2007students} and the confidence interval as having a 95\% probability of containing the population parameter \citep{andrade2016interpretation}.
Bayesian concepts, such as Bayes factor and credible intervals, are among the recommended measures of evidence as alternatives to the \(p\)-value and confidence intervals \citep{wasserstein2016}.
Therefore, exposure to and education training in Bayesian methods may help correct students' misconceptions and provide them with additional approaches to statistical inference.
Last but not least, training undergraduate students in Bayesian methods would prepare them for their prospective jobs in industry and for graduate training where they may utilize Bayesian methods.

Despite a growing literature on teaching strategies and exercises for specific Bayesian topics (\citet{RouderMorey2019TAS}, \citet{EadieHuppenkothenSpringfordMccormick2019JSE}, and \citet{BarcenaGarinMartinTusellUnzueta2019JSE} are some recent examples), little is known about the current state of undergraduate Bayesian education.
The literature suggests a limited offering of Bayesian courses at the undergraduate level.
For instance, none of the top 10 statistics programs require a Bayesian course and not all of them offer such a course \citep{hoegh2020}.
Some Bayesian educators have described their own Bayesian courses as examples \citep{Witmer2017TAS, hu2020, hoegh2020, hucontent}.
Recently, there has also been a panel of four statistics educators who teach such courses discussing their own pedagogical approaches \citep{johnson2020}.
Since the discussions around undergraduate Bayesian education are relatively new, these publications have filled a much-needed gap in the literature.
However, most of these works are based on specific courses at specific institutions.
For a broader picture of the current state of undergraduate Bayesian education, we decided to analyze Bayesian courses across different institutions.

Our primary interest in this study was to know how commonly Bayesian statistics is taught at the undergraduate level as a course and how it is taught.
For simplicity, our use of a Bayesian course refers to a Bayesian course for undergraduate students throughout the paper.
We wanted to see how Bayesian courses fit in the overall undergraduate programs.
We also wanted to have a deeper understanding of these courses, including the prerequisites required, topics covered, and the tools (e.g., software) that they adopted.
To answer these questions, we collected data from 152 high-ranking research universities and liberal arts colleges and focused on understanding whether an undergraduate Bayesian course is offered, and how each identified course fits into the undergraduate curricula.
We also collected syllabi from existing courses and conducted a series of analyses on topics covered, programming and computing tools, and assessments of these courses.

The main goals of this paper are to provide an overview of the current state of Bayesian education at the undergraduate level, provide recommendations for programs looking to offer undergraduate Bayesian courses, and share a list of resources (e.g., software and textbooks) for aspiring and current Bayesian educators.

The remainder of the paper is organized as follows. Section \ref{methods} presents the details of our data collection procedures. Our results based on analyses of program curricula and course syllabi are reported in Section \ref{results}, followed by our recommendations in Section \ref{recommendations}. We end the paper with a few concluding remarks in Section \ref{conclusion}.

\hypertarget{methods}{%
\section{Data Collection}\label{methods}}

To find out whether Bayesian courses are commonly offered, we surveyed all universities with a ranking of 100 or higher (i.e., better ranking) and liberal arts colleges with a ranking of 50 or higher based on US News rankings (\citeyear{usnews-uni}; \citeyear{usnews-college}).
To avoid confusion, we use the word ``university'' when referring to a research university, ``college'' when referring to a liberal arts college, and ``institution'' when referring to a research university or a liberal arts college.
We purposefully selected these institutions because we wanted a pool of institutions that are likely to offer Bayesian courses so that we could examine their courses in depth.
We identified 102 research universities (due to ties) and 50 colleges for further analysis.

After obtaining this sample of institutions, we collected data from the institutions' websites regarding the courses and programs they offer\footnote{Data collection was completed by the authors and an undergraduate researcher.}.
We searched the word ``Bayesian'' in course catalogs spanning two academic years, from Fall 2019 to Summer 2021.
We deliberately narrowed our search to a two-year time frame because a Bayesian course being offered in longer intervals would be very unlikely to be accessible to all students in an academic program.
In other words, while junior and senior students are the ones who are most likely to satisfy the course prerequisites, they would not have access to a Bayesian course if it is offered less frequently.
We only tracked courses that contain the word ``Bayesian'' in the title of the course and eliminated any courses that only include ``Bayesian'' in the course description.

To get an overview of how these Bayesian courses fit in the undergraduate curricula, we tracked in which department the Bayesian course is offered, whether the course is cross-listed for undergraduate and graduate enrollment, and whether it is part of any major as an elective or a required course.
To examine course prerequisites, we recorded whether the course has calculus, probability, linear algebra, statistics, and programming courses as prerequisites and recorded any other less common prerequisites we encountered.

During data collection from institutions' websites, we faced two challenges.
The first one related to how to identify whether the Bayesian course is required or an elective course for a major, since such information may not be readily available on course catalogs.
As a solution, we searched through the list of majors of an institution by looking for majors that may contain the strings of ``statistic'', ``math'', ``comput''\footnote{This is intentional to capture strings such as "computer science" or "computational science".} or ``data.''
For each major satisfying these criteria, we noted whether the identified Bayesian course is part of the major or not.
To make sure not to exclude any majors that might have possibly fallen out from our search strings (e.g., psychology), we ran a Google search with the institution name, specific course number, and the string ``major requirements.''

Our second challenge was about understanding the prerequisite requirements for the identified Bayesian courses.
First, the prerequisites and how they are offered vary from one institution to another.
Second, a prerequisite course can have its own set of prerequisites.
Moreover, at many institutions, students can take different paths leading to the Bayesian course.
Last but not least, some institutions have concurrent requirements for students enrolled in the Bayesian course.
As a solution, we assumed a hypothetical student with no prior calculus, statistics, or computer science training in high school and counted the minimum number of courses this student has to take before or concurrently enrolling in the Bayesian course.
In other words, we counted prerequisites of prerequisites in our data collection and subsequent analysis.

To examine content of courses, we relied on course syllabi.
During data collection from institutions' websites, we downloaded the available syllabi directly from the course catalogs.
We also searched department websites and conducted a Google search to locate additional syllabi online.
In addition, we reached out to the instructors who offered the course most recently and asked for their syllabi.
In total, we collected 29 syllabi.

\hypertarget{results}{%
\section{Program and Syllabi Analyses}\label{results}}

\hypertarget{results-program}{%
\subsection{Bayesian Courses in Program Curricula}\label{results-program}}

Among the 152 high-ranking institutions, we identified 46 institutions that offer a Bayesian course.
In total, we have identified 51 Bayesian courses and note that 5 universities have two Bayesian courses.
Breaking down by institution type, it is 12\% of the colleges (6 out of 50) and 39\% of the universities (40 out of 102), respectively.
Of the 45 courses offered at universities, 60\% are cross-listed between undergraduate and graduate programs, meaning that undergraduate students take these courses along with graduate-level peers.
In this section, we investigate the major disciplines and the prerequisite requirements of 51 Bayesian courses across institutions.

\hypertarget{the-breakdown-of-major-disciplines-of-the-bayesian-courses}{%
\subsubsection{The breakdown of major disciplines of the Bayesian courses}\label{the-breakdown-of-major-disciplines-of-the-bayesian-courses}}

There exist differences in how the majors are named across institutions.
To account for these differences, we adopted broad titles of majors.
For example, under ``Statistical Sciences'', we included common majors, such as ``Statistical Science'', ``Statistics'', and ``Applied Statistics''.
In this group, we also included less common majors, such as ``Actuarial Science'' and ``Biometry and Statistics''.
Moreover, we created ``Combination of Statistical, Mathematical, Computer, or Data Sciences'' to include majors such as ``Statistics and Data Science'', ``Math and Statistics'', and ``Mathematics and Computer Science'', as these names usually refer to more than one single discipline.
We summarize the major disciplines that have the Bayesian course as an elective or a required course in Table \ref{tab:major-table}.
Some Bayesian courses count towards multiple majors, thus the total number of majors exceed the total number of the identified Bayesian courses.

\begin{table}[!h]

\caption{\label{tab:major-table}Summary of major disciplines that  explicitly include any of the 51 identified Bayesian courses.}
\centering
\fontsize{10}{12}\selectfont
\begin{tabular}[t]{lrrr}
\toprule
Major Discipline & Elective & Required & Total\\
\midrule
Statistical Sciences & 29 & 2 & 31\\
Mathematical Sciences & 13 & 0 & 13\\
Combination of Statistical, Mathematical, Computer, or Data Sciences & 12 & 0 & 12\\
Data Sciences & 6 & 2 & 8\\
Computer Sciences & 5 & 0 & 5\\
Biological Sciences & 5 & 0 & 5\\
Quantitative Economics & 4 & 0 & 4\\
Business, Economics, and Management & 3 & 0 & 3\\
Psychology and Cognitive Sciences & 3 & 0 & 3\\
Public Policy and Political Science & 2 & 0 & 2\\
Others & 5 & 0 & 5\\
\textbf{Total} & \textbf{87} & \textbf{4} & \textbf{91}\\
\bottomrule
\multicolumn{4}{l}{\textsuperscript{*} The Others category includes Geological and Planetary Sciences, Quantitative Sciences,}\\
\multicolumn{4}{l}{Physics, Philosophy, and No Specific Major, each of which has one elective course.}\\
\end{tabular}
\end{table}

We observe that the Bayesian courses are offered as part of a variety of majors across institutions.
As anticipated, the Bayesian courses are often part of majors that include statistical, mathematical, computer, or data sciences, or a combination of these fields.
Although a few in number, majors in biological sciences, business, economics, and management also have a Bayesian course.
Even fewer in number, we identified Bayesian courses that are part of psychology and cognitive sciences, public policy and political science, geological and planetary sciences, philosophy, physics, and quantitative science majors.
Despite the fact that for some of these fields we only identified one instance of a Bayesian course being counted towards the major, seeing a Bayesian course as part of a variety of disciplines at the undergraduate level is extremely important and a new finding for us.
This finding sheds light on the growing importance of Bayesian methodology across these disciplines.
Moreover, it provides evidence that in addition to statistics, many disciplines are seeing value in teaching Bayesian methods to their students.

Regardless of the majors, among the 51 identified Bayesian courses, 51\% are offered in statistics departments, 8\% in computer science departments, and 2\% in math departments.
There are also many departments which are at the intersection of statistics, mathematics, computer science, and/or data science.
These departments offer 18\% of the identified Bayesian courses.
The remaining 22\% courses are taught in various departments such as physics, psychology, ecology, evolution and marine biology and two of which were cross-listed between multiple departments.

We also note that the Bayesian courses are much more often to be an elective than being a required course. Among the four instances of required courses, two are from statistical sciences while the other two are from data sciences.
One important takeaway for advancing Bayesian education at the undergraduate level, which leads to one of our recommendations in Section \ref{recommendations}, is to make Bayesian courses be part of majors and if possible, required.

\hypertarget{prerequisites}{%
\subsubsection{Prerequisite requirements of the Bayesian courses}\label{prerequisites}}

\begin{figure}

{\centering \includegraphics[width=0.7\linewidth]{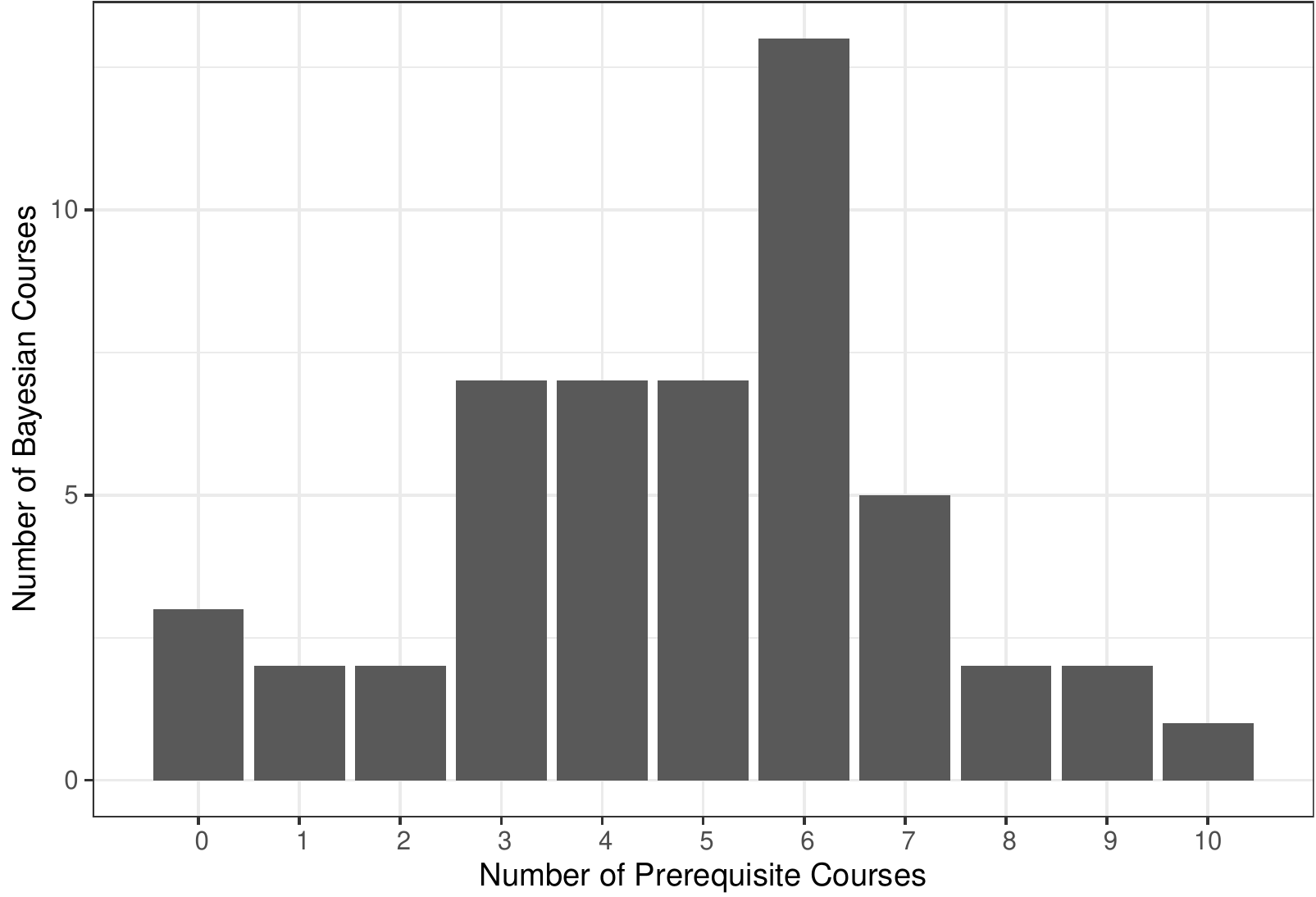} 

}

\caption{Summary of prerequisite courses of the Bayesian courses. The number includes prerequisites of prerequisites.}\label{fig:reqs-summary}
\end{figure}

An important aspect of understanding how the Bayesian courses fit in the program curricula is about what kind of preparation students need, i.e., the prerequisite requirements.
As discussed in Section \ref{methods}, we focused on the minimum number of prerequisite courses a student needs to take and counted all prerequisites of prerequisites.
We summarize the number of prerequisites in Figure \ref{fig:reqs-summary}.

As can be seen in Figure \ref{fig:reqs-summary}, the number of prerequisites ranges from zero to ten.
A majority of the Bayesian courses have 3 - 6 prerequisites, with 6 as the mode.
While we were not able to identify the class level of the courses, we conjecture that they are mostly junior/senior level courses for undergraduates given the numbers of prerequisites we identified.
It is also worth recalling that of the 45 universities offering Bayesian courses, 60\% are cross-listed between undergraduate and graduate programs.

We have also conducted an in-depth analysis of the types of prerequisite courses and their popularity. Specifically, we tracked calculus, linear algebra, computing, probability and statistics, and other courses that fall outside of this classification.
Clearly, different programs have different prior knowledge expectations from students entering the Bayesian course, leading to our conjecture that the teaching approaches can be quite different.
As a side note, we identified Bayesian courses that are offered outside of statistics departments and have no probability or statistics prerequisites, a finding that may surprise many statistics educators.

Despite the high variance in prior knowledge expectations of the Bayesian courses, there are a few conclusions we can draw.
First, undergraduate Bayesian courses are commonly calculus-based.
Second, linear algebra is a requirement mainly for courses that have high a number of prerequisites (greater than six).
Last but not least, almost all of the Bayesian courses have either a statistics, probability or statistics and probability course prerequisite.

In terms of mathematical preparation, 43\% of the 51 Bayesian courses have three calculus courses as prerequisites, 29\% require two, 10\% require only one, and 16\% do not require calculus courses at all.
Compared to calculus, linear algebra is a less common prerequisite, with only 37\% requiring it.

As computing is an important and interwoven component in a Bayesian course, we dive deeper into analyzing the computing prerequisites.
While 33\% of the 51 Bayesian courses mention some sort of computing prerequisite,
not all of these mentions are prerequisite courses.
The explicitly stated computing prerequisite courses include introductory computer programming courses, statistical computing, R for data science, data science with R, R programming, data structures, computational thinking and doing, and SAS programming.
We also identified one institution that has two computing prerequisites including an introductory computer programming course and a data structures course.
In addition, some course descriptions mention specific computing software, knowledge or skill as opposed to a specific course\footnote{For instance, some mentions include ``R recommended", ``familiarity with R", ``basics of R programming required", ``some acquaintance with fundamentals of computer programming", and ``familiarity with some programming language or numerical computing environment."}.

\begin{table}[!h]

\begin{threeparttable}
\caption{\label{tab:prob-stat-table}Summary of statistics and probability prerequisite courses of the 51 Bayesian courses.}
\centering
\fontsize{10}{12}\selectfont
\begin{tabular}[t]{>{\raggedright\arraybackslash}p{20em}>{\raggedleft\arraybackslash}p{20em}}
\toprule
Prerequisite Course & Count\\
\midrule
Probability & 16\\
Linear Models & 13\\
Probability and Statistics & 13\\
Mathematical Statistics & 11\\
Statistics & 8\\
Statistical Inference & 7\\
Statistical Methods & 7\\
Introduction to Statistics & 6\\
Bayesian Statistics & 2\\
Machine Learning & 2\\
Others & 5\\
\bottomrule
\end{tabular}
\begin{tablenotes}
\small
\item [*] The Others category includes Data Analysis and Statistical Inference, Economometrics, Foundation of Information and Inference, Introduction to Statistical Theory, and Linear Algebra, Probability, and Statistics for the Life Sciences, each of which has one occurrence.
\end{tablenotes}
\end{threeparttable}
\end{table}

Lastly, we examined the probability and statistics prerequisites in detail.
The detailed course title and the number of Bayesian courses requiring it are summarized in Table \ref{tab:prob-stat-table}.
Recall that prerequisites of prerequisites were counted in this analysis.
The median number of probability and statistics prerequisites is 2 and the range is from 0 to 5.
As can be seen, the probability and statistics prerequisite requirements vary from one course to another.
The most notable takeaway is that many programs require a course on probability and the most common statistical prerequisite is a course on linear models.

\hypertarget{results-syllabi}{%
\subsection{Syllabi Analyses}\label{results-syllabi}}

Analyzing collected syllabi can provide further insight on course content, tools, and assessments.
Among the collected 29 syllabi, there are 3 syllabi from colleges and 26 from universities.
Depending on the institution's academic calendar, these courses run between 10 weeks (quarter system) and 16 weeks (semester system) of instruction period.
Some have designated lab sessions and the majority of the courses at universities have teaching support such as one or more teaching assistants.
In our syllabi analyses, we focus on two aspects: topics covered in the course and computing tools.
A short analysis on course assessments is included in the Supplementary Materials for further reading.

\hypertarget{topic}{%
\subsubsection{Analysis of topics by area}\label{topic}}

First, we present the analysis of topics, where we group the identified topics into three general areas: foundations of Bayesian inference, Bayesian computing, and Bayesian modeling.
When discussing a topic, we include the fraction of courses (out of the total 29 collected syllabi) in the parenthesis to indicate its popularity among instructors.
We note that some syllabi are more descriptive in the topics they cover while others mention less information.
Nevertheless, we believe the identified topics and their popularity show implications of what is being considered as important by instructors when teaching Bayesian statistics to the undergraduates.

\textbf{Area 1: Foundations of Bayesian inference}

When introducing the foundations of Bayesian inference, among the single-parameter models, beta-binomial (34\%) and normal-normal (28\%) are the most popular topics chosen by the instructors.
Only a few would continue to discuss the multi-parameter model of normal (7\%).
It is worth noting that the topic of multivariate normal model, which is relatively common in graduate level courses (e.g., \citet{Hoff2009}), is rarely mentioned in the collected syllabi (3\%).
This finding suggests that instructors for undergraduate Bayesian courses, not surprisingly, tend to choose simpler problem settings to introduce Bayesian inference basics, which we believe could achieve the learning outcomes just as well.
Under these simple single-parameter settings, many instructors discuss the important concepts of conjugacy (24\%) and posterior prediction (17\%), while a few cover hypothesis testing (10\%), the sequential update nature of Bayesian inference (7\%), credible intervals (7\%), and posterior predictive checks (3\%).
The top 4 most popular topics of foundations of Bayesian inference and their popularity are presented in Table \ref{tab:topic-popularity}.

\textbf{Area 2: Bayesian computing}

Moving to the set of Bayesian computing topics, we found about 1/5 of all instructors first introduce Monte Carlo simulation (21\%).
A majority of the instructors discuss Markov chain Monte Carlo (MCMC) specifically (52\%).
Among them, about a half also discuss MCMC diagnostics (24\%).
The Gibbs sampler is the most popular MCMC algorithm (28\%), indicating that instructors place a big emphasis on posterior derivation in their Bayesian courses.
The Metropolis-Hastings algorithm (21\%) is slightly more popular than the Metropolis algorithm (17\%).
Since the Metropolis-Hastings builds on the Metropolis, it is more challenging but also more widely used.
The results suggest that instructors choose to cover MCMC algorithms that are more useful in practice.

We also note that as the popularity of Stan grows (\citet{Stan}), some instructors introduce Hamilton Monte Carlo (14\%) as an additional Bayesian computing technique.
Analyses about these additional and less common topics will be presented at the end of this section.
The top 4 most popular topics of Bayesian computing and their popularity are presented in Table \ref{tab:topic-popularity}.

\textbf{Area 3: Bayesian modeling}

When covering Bayesian modeling techniques, Bayesian linear regression is the most selected topic (69\%), immediately followed by hierarchical modeling (62\%), which encompasses many appealing features of Bayesian statistics.
Logistic regression (21\%), or more generally, generalized linear models (28\%), are also covered by many instructors.
Moreover, mixture modeling, a traditionally more advanced modeling technique, is selected by a not-so-small number of instructors (17\%), whose implementation we believe is made much simpler by various MCMC software.
The top 4 most popular topics of Bayesian modeling and their popularity are presented in Table \ref{tab:topic-popularity}.

\begin{table}[!h]

\caption{\label{tab:topic-popularity}The top 4 topics and their popularity in each of the three areas. The proportions in the Popularity column are based on 29 syllabi.}
\centering
\fontsize{10}{12}\selectfont
\begin{tabular}[t]{lll}
\toprule
Area & Topic & Popularity\\
\midrule
Foundation of Bayesian Inference & Beta-binomial & 0.34\\
 & Normal-normal & 0.28\\
 & Conjugacy & 0.24\\
 & Prediction & 0.17\\
\midrule
Bayesian computing & MCMC & 0.52\\
 & Gibbs sampler & 0.28\\
 & MCMC diagnostics & 0.24\\
 & Metropolis-Hastings & 0.21\\
\midrule
Bayesian modeling & Linear regression & 0.69\\
 & Hierarchical modeling & 0.62\\
 & GLM & 0.28\\
 & Logistic regression & 0.21\\
\bottomrule
\end{tabular}
\end{table}

\textbf{Additional topics}

The analysis of additional topics shows that many instructors more formally introduce how to choose Bayesian models, which we consider as part of the area of foundations of Bayesian inference.
Among them, model checking (14\%), model comparison (21\%), and model selection (17\%) are relatively popular topics.
We also highlight the choice of discussing more advanced priors (17\%), and even the Dirichlet process prior (3\%).
When it comes to Bayesian computing, in addition to Hamiltonian Monte Carlo (14\%), variational inferences/variational Bayes is another quite advanced topic chosen by instructors for the undergraduates (7\%).
For additional Bayesian modeling techniques, we underline the choice of missing data imputation (10\%), a topic we believe that the Bayesian approach is so natural by treating the missing data as parameters, leading to imputation steps embedded in the MCMC estimation process (\citet{rubin1987}).
The topic of Gaussian processes is favored by several instructors (10\%).
In a few relatively more advanced undergraduate courses, frequentist properties of Bayesian methods are also covered (10\%).
The remaining and less popular additional topics are included in the Supplementary Materials for further reading.

Lastly, we notice some courses choose to compare and contrast Bayesian methods to frequentist methods (17\%), a decision we believe is made mostly based on prerequisites and student background.

\hypertarget{programming}{%
\subsubsection{Analysis of adopted programming and computing tools}\label{programming}}

Given the important role that computing plays in the undergraduate Bayesian courses, we report our analysis of the programming language(s), computing methods for MCMC estimation, and additional programming packages highlighted by some instructors.
Out of the 29 syllabi, 27 contain information about the programming language(s) used in the course.
R is the most popular single programming language (70\%), while Python is the only other choice (11\%).
The remaining instructors either require two languages: R and Python (4\%) or R and SAS (4\%); or they allow students to choose from a selection: R or Python (7\%), or R or Python or Julia (4\%).

When it comes to software choice, 17 courses explicitly discuss the MCMC estimation software they utilize.
Stan is the most popular single programming language (41\%), with JAGS as the second (35\%).
There are also a number of instructors who choose to introduce both (24\%). One course that utilizes both R and SAS programming languages also uses PROC MCMC which is the procedure for fitting Bayesian models in SAS.

Both JAGS and Stan can be used within R with several R packages (also available for Python).
Students in these courses would learn and practice writing JAGS and/or Stan scripts which are descriptive of the Bayesian models.
However, we note that among the 7 courses introducing Stan, at least 1 syllabus mentions their use of the brms package (\citet{brms})\footnote{This package includes a variety of wrapper functions of Stan, which ask for user's inputs such as the model formula and prior choices. 
Compared to writing JAGS or Stan scripts, these wrapper functions overall mimic the Stan script in R syntax. For example, fitting a Bayesian linear regression with the brms package requires inputs similar to those in the \texttt{lm()} R function.
For those proficient in R, these wrapper functions make it possible to fit Bayesian models without having to learn additional syntax.}.
Therefore, we cannot be certain whether each of the 7 courses introducing Stan relies on packages such as brms or actually requires writing Stan scripts.
On a related note, 1 course specifically provides no mention of Stan but highlights the use of the brms package and another Stan-based wrapper package called the rstanarm (\citet{rstanarm}).

However, the above analysis does not claim that not all instructors choose MCMC software for MCMC estimation when teaching undergraduate Bayesian courses.
In fact, 1 of the 4 courses which use both JAGS and Stan explicitly states that ``The use of software platforms (`blackbox'), such as `jags', `bugs', and `stan', is permitted only when specified by the instructions.''
We conjecture that this might not be the only course among our collected syllabi that focuses only self-coded MCMC algorithms.
In fact, our analysis of Bayesian computing topics in Section \ref{topic} reveals that many courses do introduce MCMC algorithms by self-coding.

Given the fact that 59\% of our collected syllabi introduce MCMC estimation software, we believe that at the undergraduate level, instructors prefer MCMC software over self-coded MCMC algorithms.
This software-oriented approach undoubtedly broadens the scope of Bayesian modeling techniques that can be covered in an undergraduate Bayesian course, while still maintaining a reasonably high intensity of Bayesian computing, especially if writing JAGS and/or Stan scripts is expected.

\hypertarget{recommendations}{%
\section{Future of Undergraduate Bayesian Education}\label{recommendations}}

Given our findings in Section \ref{results}, we share our vision for the future of undergraduate Bayesian education.
We provide three broad recommendations: two at the program level and one at the course level.
An additional recommendation on using a variety of assessments is included in the Supplementary Materials for further reading.
We then share sample weekly schedules for potential Bayesian courses and a list of required and recommended textbooks from our syllabi analysis.

\hypertarget{recommendations-1}{%
\subsection{Recommendations}\label{recommendations-1}}

~~~~~\textbf{Recommendation 1: Expand the access to Bayesian courses}

With Bayesian statistics becoming ever more popular and with the overabundance of data, we expect a higher demand for Bayesian statisticians and data scientists in the workforce and in research.
More students, especially those in statistical, data, mathematical, and computer sciences, will need to have access to Bayesian courses in the (near) future.
To make Bayesian courses more widely available, we recommend the following for undergraduate programs.

\begin{enumerate}
\def\labelenumi{\arabic{enumi}.}
\item
  Offer an undergraduate course in Bayesian statistics.
  While this may not be an option for many programs for various reasons, we believe offering a course to students may come in forms different from creating a new course.
  For example, institutions can rely on course-sharing infrastructures from nearby institutions\footnote{As an example, Harvey Mudd College's Bayesian Statistics course is shared among the nearby 5 colleges.}, online course sharing opportunities \citep{Hu2019CHANCE}\footnote{As an example, Vassar College's Bayesian Statistics is shared on a 10-college online consortium, Liberal Arts Collaborative for Digital Innovation (LACOL). For more information, visit \url{https://lacol.net/category/collaborations/course-sharing/}.}, or credit transfer opportunities.
  Undergraduate programs should actively search for and create such opportunities, as well as provide students with guidance.
\item
  Reduce the number of prerequisites.
  Statistics educators have been suggesting flattening prerequisites in the statistics curriculum for a while \citep{Cobb2015TAS} and we believe this applies to Bayesian courses as well.
  As our study findings have shown, the number of prerequisites for Bayesian courses varies greatly.
  To make the course accessible to a larger student body, programs need to consider reducing the number of prerequisites.
  Section \ref{prerequisites} shows many Bayesian courses with fewer prerequisites.
  We recommend core prerequisites of probability, multivariable calculus, and a statistics course (linear models recommended).
  Statistical computing is also an essential prerequisite of a Bayesian course.
  Rather than requiring a full course on computing, programs may consider integrating computing into earlier statistics courses consistent with the recommendations for such courses \citep{nolan2010, Horton2015a, Horton2015b}.
\item
  Include Bayesian modules as part of existing courses.
  Programs that are not able to offer a full course on Bayesian statistics can consider including Bayesian topics as part of other courses. With sufficient motivation and foundation building, coupled with appropriate computing tools, a Bayesian module could be introduced at statistics courses at all levels, from the introductory to the advanced \citep{hoegh2020}.
\end{enumerate}

\textbf{Recommendation 2: Make Bayesian courses a part of the majors}

\begin{enumerate}
\def\labelenumi{\arabic{enumi}.}
\item
  Consider making the Bayesian course required for statistics and data science majors.
  Examples include Duke University's statistical science major and University of California, Irvine's data science major.
\item
  If the Bayesian course is an elective, then make it a highly recommended elective.
  Many students may not know of the importance of Bayesian statistics in their studies.
  Therefore, it is crucial for programs and advisers to explicitly recommend the Bayesian course.
  For instance, if a mathematics and statistics program recommends any upper-level course as an elective for the major, students may not necessarily know the difference in choosing between a Bayesian or a geometry course.
  Therefore, programs should explicitly state that the Bayesian course is highly recommended.
\item
  Consider making the Bayesian course an elective for majors beyond the statistical, mathematical, and computational sciences.
  As our findings on major disciplines in Section \ref{results-program} suggest, Bayesian courses can be, and already are, part of many programs, such as economics, psychology, business, and biology, especially for quantitative concentrations within these programs.
  Training students in Bayesian methods from a variety of fields may help with correcting misconceptions about statistical inference, such as mis- and over-use of \(p\)-values, and providing them with additional statistical inference approaches.
\end{enumerate}

\textbf{Recommendation 3: Balance statistics with computing}

The need for curricular reform in statistics to incorporate computing has been called for by many scholars and educators \citep{nolan2010}.
Our analyses in Section \ref{results} have shown that many courses embrace computing and statistics simultaneously and there are popular tools shared among instructors.
Given our findings, we have the following recommendations for balancing statistics with computing in a Bayesian course.

\begin{enumerate}
\def\labelenumi{\arabic{enumi}.}
\item
  Introduce simulation-based learning early in the course.
  Take the opportunity to compare and contrast simulation-based inference (e.g., Monte Carlo simulation) and analytical solutions when introducing conjugate models (\citet{hu2020}).
\item
  Encourage students to write self-coded MCMC algorithms for relatively simple multi-parameter models.
  Provide ample sample scripts for students to practice their statistical programming skills. Several articles discussed their strategies of writing self-coded MCMC algorithms for simple multi-parameter Bayesian models before moving to MCMC software (\citet{hu2020}, \citet{johnson2020}, \citet{alberthu2020}, \citet{hucontent}).
\item
  If the course puts equal emphasis on computing and modeling, consider adopting one of the popular probabilistic programming languages for Bayesian model estimation through MCMC (e.g., JAGS and Stan).
  Clearly illustrate the purpose of each coding component and how it connects to important MCMC topics, including MCMC diagnostics.
  Emphasize on posterior summary and inference.
  Provide ample sample scripts for students to get familiarized with the syntax.
\item
  If the course has a slightly stronger emphasis on modeling over computing, consider introducing one of the wrapper packages for Stan for its simpler posterior summary procedure (e.g., rstanarm and brms).
  The rest of the recommendation follows number 3 above.
\end{enumerate}

\hypertarget{sample-weekly-schedules-for-10-week-and-14-week-bayesian-courses}{%
\subsection{Sample weekly schedules for 10-week and 14-week Bayesian courses}\label{sample-weekly-schedules-for-10-week-and-14-week-bayesian-courses}}

To help aspiring Bayesian educators get started with creating a new course, we provide two sample schedules.
These sample schedules are designed based on the analysis of topics by areas reported in Section \ref{topic}.
Our recommended topics in the sample schedules are selected based on their popularity, summarized in Table \ref{tab:topic-popularity}.
Table \ref{tab:sample-schedule} lists the two sample schedules side by side.

\begin{table}[!h]

\caption{\label{tab:sample-schedule}Sample weekly schedules for 10- and 14-week academic terms. Numbers refer to week indexes, e.g., 1 means Week 1.}
\centering
\fontsize{10}{12}\selectfont
\begin{tabular}[t]{llll}
\toprule
Area & Topic & 10-week & 14-week\\
\midrule
Foundation of Bayesian Inference & Beta-binomial & 1 & 1\\
 & Normal-normal & - & 2\\
 & Conjugacy & 2 & 3\\
 & Prediction and predictive checks & 3 & 4\\
\midrule
Bayesian computing & MCMC & 4 & 5\\
 & Gibbs sampler & 5 & 6\\
 & Metropolis-Hastings & 6 & 7\\
 & MCMC diagnostics & 7 & 8\\
\midrule
Bayesian modeling & Linear regression & 8 & 9\\
 & Hierarchical models & 9 & 10\\
 & Logistic regression & 10 & 11\\
 & GLM & - & 12\\
\midrule
Additional topics & Model comparison and model selection & - & 13\\
 & Missing data imputation, more on priors & - & 14\\
\bottomrule
\end{tabular}
\end{table}

The sample schedule for a 10-week course spends 3 weeks on foundations of Bayesian inference, 4 weeks on Bayesian computing, and the remaining 3 weeks on Bayesian modeling.
Depending on the learning objectives, instructors can choose to shorten the number of weeks on computing and lengthen that on modeling, especially if the course has an applied focus.
One topic that requires special attention in the 10-week schedule is model evaluation (comparison and selection) which is an important aspect of Bayesian modeling.
It is especially crucial if final projects are assigned as students would need rigorous methods for finalizing their model.
Therefore, it is important that instructors emphasize model evaluation throughout the course, starting early on with predictive checks and later also utilizing predictive checks for linear regression, hierarchical, and logistic regression models.

For the sample schedule of a 14-week course, we include the area of additional topics, where the instructors can mix and match from the identified additional topics presented in Section \ref{topic}.
In the sample schedule, we include the topics of the highest popularity, and instructors should delete and add as they see fit.
For example, if students in the course have prior exposure to frequentist inference, then comparing Bayesian inference to frequentist inference for common models, such as linear regression, would undoubtedly be a suitable additional topic.
The aforementioned emphasis on model evaluation throughout the course also applies to the 14-week schedule.

\hypertarget{required-and-recommended-textbooks-from-the-syllabi}{%
\subsection{Required and recommended textbooks from the syllabi}\label{required-and-recommended-textbooks-from-the-syllabi}}

To further help aspiring Bayesian instructors to get started, Table \ref{tab:books} contains information of the required and recommended textbooks collected from the syllabi analyses.
The column ``Required'' shows the number of courses (out of 29) requiring the textbook while the column ``Recommended'' shows the number of courses recommending the textbook.
The textbooks are ranked by the ``Required'' column.

\begin{table}[H]
\centering
\caption{List of required and recommended textbooks from 29 collected syllabi.}
\label{tab:books}
\vspace{5mm}
\resizebox{\columnwidth}{!}{%
\begin{tabular}{l l r r}
\hline
Book Title & Year & Required & Recommended \\ \hline
Doing Bayesian Data Analysis: A Tutorial with R, JAGS, and Stan & \citeyear{Kruschke2014}   & 4 & 3 \\ Bayesian Data Analysis   & \citeyear{BDA3}   & 4 & 4 \\
A First Course in Bayesian Statistical Methods  & \citeyear{Hoff2009}   & 4 & 3 \\
Statistical Rethinking: A Bayesian Course with Examples in R and Stan   & \citeyear{rethinking2}    & 1 & 4 \\
Bayesian Statistical Methods    & \citeyear{BSM}    & 2 & 0 \\
Bayes Rules! An Introduction to Bayesian Modeling with R    & \citeyear{bayesrules} & 2 & 0 \\
Probability and Bayesian Modeling   & \citeyear{AlbertHu2019book}   & 1 & 1 \\
A Student's Guide to Bayesian Statistics    & \citeyear{Lambert2018book}    & 1 & 1 \\
Applied Bayesian Statistics: With R and OpenBUGS Examples   & \citeyear{ABS2013book}    & 1 & 0 \\
Data Analysis: A Bayesian Tutorial  & \citeyear{DA2006book} & 1 & 0 \\
Bayesian Ideas and Data Analysis: An Introduction for Scientists and Statisticians  & \citeyear{BIDA2010book}   & 0 & 1 \\
Bayesian Methods for Data Analysis  & \citeyear{BMDA2008book}   & 0 & 1 \\
Bayesian Computation with R & \citeyear{BCWR2009book}   & 0 & 2 \\
Introduction to Bayesian Statistics & \citeyear{IBS2016book}    & 0 &   1 \\ \hline
\end{tabular}%
}
\end{table}

\hypertarget{conclusion}{%
\section{Concluding Remarks}\label{conclusion}}

With the advances of computing and the emergence of data science, we expect the demand for statisticians and data scientists trained in Bayesian statistics to continue growing.
Undergraduate programs need to prepare to meet this demand.
In this paper, we provide an overview of current practices in undergraduate programs and Bayesian courses.
We also give several recommendations, some at the program level while others at the course level, as a guide to new courses that undergraduate programs may develop.
Moreover, two weekly sample schedules are proposed given the popularity of topics analyzed from the collected syllabi, and a list of required and recommended textbooks is provided.

Any study with data, especially the ones with data collected from college and university websites, comes with its own set of limitations, which sets the parameters of our reported results and conclusions.
First, our primary goals were identifying how common Bayesian courses are and understanding the content of these offered courses.
To achieve these goals, we avoided random sampling of institutions and instead chose to examine programs and courses at highly ranked institutions, with the assumption that they are more likely to offer Bayesian courses given their resources.
Therefore, it is possible that Bayesian courses are less common in undergraduate programs than what we have reported here.
In addition, recall that there were a number of Bayesian courses cross-listed between graduate and undergraduate programs.
Given the limited available information, it was not possible to evaluate the appropriateness of these courses specifically for undergraduate learners.

Another interest of ours was understanding whether students graduating from undergraduate programs, especially those in statistical, mathematical, computer, and data sciences have access to Bayesian courses.
We were able to count how many institutions offer these courses in the sample we have examined.
However, more students from different institutions might have access to undergraduate Bayesian courses through less common opportunities.
Although we have not formally tracked all such possibilities, we have encountered and reported several examples of course sharing infrastructures where students in one institution can take a class in another institution.
There may be more such course sharing practices that we do not know.

We were also interested in understanding to which majors Bayesian courses count towards.
One limitation in identifying majors was that some majors may count the Bayesian course towards the major but may not necessarily list it as an explicit elective.
For instance, if a quantitative economics program counts any statistics courses with a label 400 and above towards the major, we may have missed it in our study.
Such information would have been nearly impossible to track.

In addition, our main data source was institution websites including department web pages, course catalogs, and course pages.
We had to use a fair use of human judgment in understanding course titles, especially for some courses whose titles and descriptions are different from the courses that we are used to seeing in statistics programs.
Moreover, for a few courses, some information on the web could be outdated or incomplete.
For instance, when we were tracking prerequisites, only a few of the prerequisites mentioned statistical software.
However, we know from our own experience that even at the introductory level, statistical software are often adopted.
Therefore, it is possible that computing prerequisites for Bayesian courses might be more common than what is reported in our findings.

There remain unanswered questions about the current state of undergraduate Bayesian education that call for future research.
In our study, we focused on standalone Bayesian courses but Bayesian methods may be covered as part of other undergraduate courses.
For instance, a course on linear models may in fact include regression from Bayesian perspective but the course title or course descriptions may not necessarily capture the Bayesian content.
Therefore, we call for future statistics education research to capture such information in a survey format from the broader statistics community beyond Bayesian instructors.

An important note we would like to make is that we both teach our own Bayesian courses and have our own textbooks with undergraduates as the main targeted audience.
With a few years of experience in teaching the subject, we have our own views on how and what should be taught in Bayesian statistics.
However, everything we report here, including the subsequent recommendations, is based on the courses and syllabi we examined in the study.
We tried our best to remove any bias we may have from our own teaching experiences.

Last but not least, we invite current and future Bayesian educators to join the undergraduate Bayesian education network\footnote{\url{https://undergrad-bayes.netlify.app/}}, an online community that fosters discussions of undergraduate Bayesian education.

\bibliographystyle{agsm}
\bibliography{bibliography.bib}

\newpage

\hypertarget{additional-topics-from-syllabi-analyses}{%
\section*{Supp 1: Additional topics from syllabi analyses}\label{additional-topics-from-syllabi-analyses}}

We include the additional topics identified from our syllabi analysis in Table \ref{tab:additional-topics}. We group them into the three general areas: 1) Foundations of Bayesian inference, 2) Bayesian computing, and 3) Bayesian modeling. We note that some topics can be classified into more than one area.

\begin{table}[!h]

\caption{\label{tab:additional-topics}List of additional topics}
\centering
\fontsize{10}{12}\selectfont
\begin{tabular}[t]{lll}
\toprule
Foundation & Computing & Modeling\\
\midrule
Dirichlet-Multinomial & Data augmentation & Topic models\\
Gamma-Poisson & Tempering methods & Deep generative models\\
Bayes factor & Importance sampling & State space models\\
Empirical Bayes & Laplace approximation & Nonparametric models\\
AIC, DIC, WAIC & Collapsed Gibbs sampling & Semiparametric models\\
Model averaging & Blackbox variational inference & Hidden Markov models\\
Shrinkage estimators & Sequential Monte Carlo & Bayesian tree models\\
Decision theory & Elliptical slice sampling & Measurement error\\
ANOVA and Poisson ANOVA & Bayesian optimization & Mixed-effects models\\
 & Nested sampling & Multinomial models\\
 & EM & Ordinal models\\
 & Reversible jumps MCMC & Count models\\
 & Rejection sampling & Graphical models with Stan\\
 & Tensorflow, PyTorch, PyMC3 & Signal detection theory\\
 &  & Inference in language and vision\\
 &  & Classification\\
 &  & Bayesian network\\
 &  & Time series modeling\\
 &  & Cognitive modeling\\
 &  & Naïve Bayes classification\\
\bottomrule
\end{tabular}
\end{table}

\hypertarget{assessment}{%
\section*{Supp 2: Analysis of course assessments from syllabi analyses}\label{assessment}}

We believe the assessment tools and their grade breakdown highly depend on the instructor, and possibly the institution as well.
Nevertheless, analyzing this information provides useful insight into instructors' choices of course assessments.
Our analysis focuses on the types of the assessments, their popularity, and additional useful information presented in the collected syllabi. Table \ref{tab:assessment} presents details of the top 5 most popular assessment tools.

\begin{table}[H]
\centering
\caption{Details of the top 5 most popular assessment tools}
\label{tab:assessment}
\vspace{5mm}
\resizebox{\columnwidth}{!}{%
\begin{tabular}{l l l}
\hline
Assessment tool & Count & Key takeaways or examples \\ \hline
Homework & 29 & Every course assigns homework. \\
Exams & 18 & Take-home exams or a combination of take-home and in-class exams. \\ 
Projects & 18 & Focus of data analysis; various project products. \\ 
Participation & 11 & Two examples: 1. paper reading and discussions; 2. students lead one in-class discussion. \\
Quiz & 9 & Low to no-grade contribution; assigned frequently. \\ \hline
\end{tabular}%
}
\end{table}

Homework is, without doubt, the most popular assessment tool (every course has a homework component).
When it comes to exams, take-home exams seem a common choice, and we note that one instructor explicitly describes the take-home exam as ``open book open notes, including data analysis with R.''
We believe the popularity of the take-home exam choice reflects the instructors' emphasis on assessing students' ability to conduct Bayesian data analysis, which usually involves using statistical software and therefore requires the take-home format.
As for the variety of project products, formats include (a combination of) a final report, a presentation, a proposal, and a poster review.
One instructor explicitly opens up the product types to report, or poster, or blog post, or video tutorial, or Shiny apps.
These results manifest the instructors' emphasis on conducting data analysis as the core of the projects in an undergraduate Bayesian course, and their overall balancing of oral presentation and written report in communicating the results.
The 4th and the 5th most popular assessment tools are participation and quizzes, respectively.

We end our assessment analysis by listing a few unique assessment tools (separate from the aforementioned 5 common assessments): lab reports, article reflection, and case studies.
It is encouraging to see a variety of assessment tools used by instructors for their undergraduate Bayesian courses, some of which reflect the instructors' beliefs (e.g., the importance of data analysis with statistical software in take-home exams) while others propose fresh perspectives on assessments (e.g., reflection of reading articles).

\hypertarget{recommendation-4-use-a-variety-of-assessments}{%
\section*{Supp 3: Recommendation 4: Use a variety of assessments}\label{recommendation-4-use-a-variety-of-assessments}}

While some assessment tools are certainly more popular than others, our biggest takeaway from our assessment analysis in Section \ref{assessment} is to use a variety of assessments.
We present our specific recommendations below.

\begin{enumerate}
\def\labelenumi{\arabic{enumi}.}
\item
  Include a data analysis portion in a homework assignment and in an exam. This practice helps emphasize the importance of Bayesian computing and adequately assess students' learning and programming skills.
\item
  Include a course project component in course assessments. Make it open-ended and allow students to freely explore their interests. Encourage varied presentation styles, such as report, oral presentation, and video tutorial, to name a few. Provide guidance throughout the process.
\item
  Include formative assessments such as quizzes (low to no-grade contribution) to get students to do drill and practice exercises. Such short assessments provide an opportunity for retention of new terminology and getting familiar with new notation.
\item
  Be creative with assessment tools: Open-ended case studies allow students to work on new models in new contexts; reading and discussing accessible journal articles build students' confidence and enhance their ability of conducting research; computing labs allow students further practice their programming skills. Interested readers can refer to Section 3 of \citet{hu2020} for details of these approaches in an undergraduate Bayesian course.
\item
  In addition to lectures, use real data in every assessment when possible.
\end{enumerate}

\end{document}